\magnification=\magstep 1
\overfullrule=0pt
\hfuzz=16pt
\voffset=0.0 true in
\vsize=8.8 true in
\baselineskip 20pt
\parskip 6pt
\hoffset=0.1 true in
\hsize=6.3 true in
\nopagenumbers
\pageno=1
\footline={\hfil -- {\folio} -- \hfil}

\centerline{\bf Dynamics of Nonequilibrium Deposition with
Diffusional Relaxation}

\vskip 0.4in

\centerline{\bf Vladimir Privman}

\vskip 0.2in

\centerline{\sl Department of Physics, Clarkson University,
Potsdam, New York 13699--5820, USA}

\vskip 0.4in

\centerline{\bf ABSTRACT}

Models of adhesion of extended particles on linear and
planar substrates are of interest in interpreting surface
deposition in colloid, polymer, and certain biological
systems. An introduction is presented to recent theoretical
advances in modeling these processes. Effects of
diffusional relaxation are surveyed in detail, including
results obtained by analytical, large-scale numerical,
mean-field and scaling approaches.

\vfil

\noindent{}Review article for\ {\sl Annual Reviews in Computational
Physics}.

\eject

\noindent{\bf 1. Introduction}

\

Dynamics of important physical, chemical, and biological
processes, which have received attention recently [1],
provides examples of strongly fluctuating systems in low
dimensions, $D=1$ or 2. These processes include surface
adsorption, for instance of colloid particles or proteins,
possibly accompanied by diffusional or other
relaxation, for which the experimentally relevant dimension
is that of planar substrates, $D=2$. For reaction-diffusion
kinetics in chemistry, the classical studies were for
$D=3$. However, recent emphasis on heterogeneous catalysis
sparked interest in $D=2$. In fact for both deposition
and reactions, some experimental results exist even in
$D=1$ (literature citation of some specific results
will be given
later). Finally, kinetics of ordering and phase separation,
largely amenable to experimental probe in $D=3$ (with
fewer results available in $2D$), attracted much
theoretical effort in $D=1,2$.

Recent theoretical emphasis on low-dimensional models has
been driven by the following interesting combination of
properties. Firstly, as in most branches of theoretical
sciences, models in $D=1$ and occasionally, in $D=2$, allow
derivation of analytical results. Secondly, it turns out
that all three types of models: deposition-relaxation,
reaction-diffusion, phase separation, are interrelated in
many, but not all, of their properties. This observation is
by no means obvious, and in fact it is model-dependent and
can be firmly established and explored only in low
dimensions, especially in $D=1$\ [1].

Lastly, it turns out that for such systems with stochastic
dynamics without the equilibrium state, important regimes
(for instance, the large-time asymptotic behavior) are
frequently governed by strong fluctuations manifesting in
power-law rather than exponential time dependence, etc.
However, the upper critical dimension above which the
fluctuation behavior is described by the mean-field
(rate-equation) approximation, is typically lower than in
the more familiar and better studied equilibrium
phase transition models. As a result, attention has been
drawn to low dimensions where the strongly fluctuating
non-mean-field behavior can be studied.

Low-dimensional nonequilibrium dynamical models pose
several interesting challenges theoretically and
numerically. While many exact, asymptotic, and numerical
results are already available in the literature [1], this
field presently provides examples of properties (such as
power-law exponents) which lack theoretical explanation
even in $1D$. Numerical simulations are challenging and
require large scale computational effort already for $1D$
models. For more experimentally relevant $2D$ cases, where
analytical results are scarce, difficulty in numerical
investigations has been the ``bottle neck'' for
understanding many open problems.

The purpose of this review is to provide an introduction to
the field of nonequilibrium surface deposition models of
extended particles. No comprehensive survey of the
literature is attempted. Relation of deposition to other
low-dimensional models mentioned earlier will be only
referred to in detail in few cases. The specific models
and examples selected for a more detailed exposition, i.e.,
models of deposition with diffusional relaxation, were
biased by author's own recent work.

The outline of the review is as follows. The rest of this
introductory section is devoted to defining the specific
topics of surface deposition to be surveyed. Section~2
describes the simplest models of random sequential
adsorption. Section~3 is devoted to deposition with
relaxation, with general remarks followed by definition of
the simplest, $1D$ models of diffusional relaxation for
which we present a more detailed description of various
recent theoretical results. Multilayer deposition is also
addressed in Section~3. More numerically-based $2D$ results
for deposition with diffusional relaxation are presented in
Section~4, along with concluding remarks.

Surface deposition is a vast field of study. Indeed,
dynamics of the deposition process is governed by substrate
structure, substrate-particle interactions,
particle-particle interactions, and transport mechanism of
particles to the surface. Furthermore, deposition
processes may be accompanied by particle motion on the
surface and by detachment. Our emphasis here will be on
those deposition processes where the particles are
``large'' as compared to the underlying atomic and
morphological structure of the substrate and as compared to
the range of the interparticle and particle-substrate
interactions. Extensive theoretical study of such systems
is relatively recent and it has been motivated by
experiments where submicron-size colloid, polymer, and
protein ``particles'' were the deposited objects [2-17].
Indeed, the submicron sizes are about two orders of
magnitude larger than the atomic dimensions and about one
order of magnitude larger than the range of the typical
double-layer and Van der Waals interactions. In contrast,
adhesion processes associated, for instance, with crystal
growth [18], involve atomic-size interactions and while the
particle-particle exclusion is always an important factor,
its interplay with other processes which affect the growth
dynamics is quite different.

Thus we assume that the main mechanism by which particles
``talk'' to each other is exclusion effect due to their
size. Perhaps the simplest and the most studied model with
this feature is Random Sequential Adsorption (RSA). The RSA
model, to be described in detail in Section~2, assumes that
the particle transport to the surface is summarized by a
uniform deposition attempt rate $R$ per unit time and area,
due to the incoming particle flux. The effect of interactions
is simplified to allow only monolayer deposition
(presumably mimicing repulsive particle-particle and
attractive particle-substrate forces). Within this
monolayer deposit, each new arriving particle must either
``fit in'' in an empty area allowed by the hard-core
exclusion interaction with the particles deposited earlier,
or the deposition attempt is rejected.

As mentioned, the basic RSA model will be described first
(Section~2). However, recent work has been focused on its
extensions to allow for particle relaxation by diffusion
(Sections~3 and 4), to include detachment processes, and to
allow multilayer formation. The latter two extensions will
be surveyed in Section~3. Our detailed presentation will be
focussed on diffusional relaxation. Of course, many other
extensions will not be discussed, such as for instance
``softening'' the hard-core interactions [12,19] or modifying
the particle transport mechanism, etc.~[20-21].

\

\noindent{\bf 2. Random Sequential Adsorption}

\

The basic irreversible RSA process [20-21] to model
experiments of colloid particle deposition [3-15] would
assume a planar $2D$ substrate and, in the simplest case,
continuum (off-lattice) deposition of spherical particles.
However, other RSA models have received attention. Indeed,
in $2D$, noncircular cross-section shapes as well as various
lattice-deposition models were considered [20-21].
Several experiments on polymers [2] and
attachment of fluorescent units on DNA
molecules [17] (the latter, in fact, is usually accompanied
by motion of these units on the DNA ``substrate'' and
detachment) suggest consideration of the lattice-substrate
RSA processes, in $1D$. RSA processes have also found
applications in traffic problems and certain other fields
and they were reviewed extensively in the literature [20-21].
Thus our presentation in this section will be limited to
defining few RSA models and outlining characteristic
features of their dynamics.

Figure~1 illustrates the simplest possible monolayer
lattice RSA model: irreversible deposition of dimers on
the linear lattice. An arriving dimer, illustrated by $a$
in the figure, will be only deposited if the underlying
pair of lattice sites are both empty. Thus, the deposition
attempt $a$ as shown in Figure~1 will succeed. However, if
the arriving particle were at $c$ or $d$ (see Figure~1), it
would be discarded. Initially, the substrate is usually
assumed to be empty. In the course of time $t$, the
coverage, $\rho (t)$, increases and builds up to order 1 on
the time scales of order $\left( R V \right)^{-1}$, where
$R$ was defined earlier as the deposition attempt rate per
unit time and ``area'' of the $D$-dimensional surface,
while $V$ is the particle volume.

At large times the coverage approaches the jammed-state
value where only gaps smaller than the particle size were
left in the monolayer. The resulting state is less dense
than the fully ordered ``crystalline'' coverage. For the
$D=1$ deposition shown in Figure~1 the fully ordered state
would have $\rho =1$. The variation of the RSA coverage is
illustrated by the solid line in Figure~2.

At early times the monolayer deposit is not dense and the
deposition process is largely uncorrelated. In this regime,
mean-field like low-density approximation schemes are
useful [22-25]. Deposition of $k$-mer particles on the linear
lattice in $1D$ was in fact solved exactly for all times
[2,26-27]. In $D=2$, extensive
numerical studies were reported [25,28-39] of the variation of
coverage with time and large-time asymptotic behavior
which is discussed in the next two paragraphs. Some exact
results for correlation properties are also available,
in $1D$ [26].

The large-time deposit has several characteristic properties
which attracted much theoretical interest. For lattice
models, the approach to the jammed-state coverage is
exponential [39-41]. This was shown to follow from the
property that the final stages of deposition are in few
sparse, well separated surviving ``landing sites.''
Estimates of decrease in their density at late stages
suggest that

$$ \rho(\infty ) - \rho(t) \sim \exp \left( -R \ell^D t
\right) \;\; , \eqno(2.1) $$

\noindent{}where $\ell$ is the lattice spacing. The
coefficient in (2.1) is of order $\ell^D/V$ if the coverage
is defined as the fraction of lattice units covered, i.e.,
the dimensionless fraction of area covered, also termed the
coverage fraction, so that coverage as density of particles
per unit volume would be $V^{-1} \rho$. The detailed
behavior depends of the size and shape of the depositing
particles as compared to the underlying lattice unit
cells.

However, for continuum off-lattice deposition, formally
obtained as the limit $\ell \to 0$, the approach to the
jamming coverage is power-law. This interesting behavior
[40-41] is due to the fact that for large times the remaining
voids accessible to particle deposition can be of sizes
arbitrarily close to those of the depositing particles.
Such voids are thus reached with low probability by the
depositing particles the flux of which is uniformly
distributed. The resulting power-law behavior depends on
the dimensionality and particle shape. For instance, for
$D$-dimensional cubes of volume $V$,

$$ \rho(\infty ) - \rho(t) \sim { \left[\ln ( RVt )
\right]^{D-1} \over RVt } \;\; , \eqno(2.2) $$

\noindent{}while for spherical particles,

$$ \rho(\infty ) - \rho(t) \sim ( RVt )^{-1/D} \;\; .
\eqno(2.3) $$

\noindent{}For the linear surface, the $D=1$ cubes and
spheres both reduce to the deposition process of segments
of length $V$. This $1D$ process is exactly solvable [26].

The $D>1$ expressions (2.2)-(2.3), and similar relations
for other particle shapes, etc., are actually empirical
asymptotic laws which have been verified for $D=2$
by extensive numerical simulations
[28-39]. The most studied $2D$ geometries are circles
(corresponding also to the deposition of spheres on the
plane where the ``$2D$-spherical'' particles are the $2D$
cross-sections) and squares. The jamming coverages are
[28-30,38-39] 

$$ \rho_{\rm squares}
(\infty) \simeq 0.5620 \;\;\;\;\; {\rm and}
\;\;\;\;\; \rho_{\rm circles}(\infty) \simeq 0.544 
\; {\rm to} \; 0.550 \;\; . \eqno(2.4) $$

\noindent{}For square particles, the crossover to continuum
in the limit $k \to \infty$ and $\ell \to 0$, with fixed
$V^{1/D}=k\ell$ in deposition of $k \times k \times \ldots
\times k$ lattice squares, has been investigated in some
detail [39], both analytically (in any $D$) and numerically
(in $2D$).

The correlations in the large-time ``jammed'' state are
different from those of the equilibrium random ``gas'' of
particles with density near $\rho (\infty )$. In fact, the
two-particle correlations in continuum deposition develop a
weak singularity at contact, and correlations generally
reflect the infinite memory (full irreversibility) of the
RSA process [26,30,41].

\

\noindent{\bf 3.~Deposition with Relaxation}

\

Monolayer deposits may ``relax'' (i.e., explore more
configurations) by particle motion on the
surface and by their detachment. In fact,
detachment has been
experimentally observed in deposition of colloid particles
which were otherwise quite immobile on the surface [6].
Theoretical interpretation of colloid particle detachment data
has proved difficult, however, because binding to the substrate
once deposited, can be different for different particles
(while the transport to the substrate, i.e., the flux of
the arriving particles in the deposition part of the
process, typically by convective diffusion, is more
uniform). Detachment also plays role in deposition on DNA
molecules [17]. Theoretical interpretation of the latter
data, which also involves hopping motion on DNA, was
achieved by mean-field type modeling [42].

Recently, more theoretically motivated studies of the
detachment relaxation processes, in some instances with
surface diffusion allowed as well, have lead to interesting
models [43-49]. These investigations did not always assume
detachment of the original units. For instance, in the $1D$
dimer deposition shown in Figure~1, each dimer on the
surface could detach and open up a ``landing site'' for
future deposition. However, in order to allow deposition
in the location represented schematically by the dimer
particle $d$, two monomers must detach (marked by an arrow)
which were parts of different dimers. Such models of
``recombination'' prior to detachment, of $k$-mers in
$D=1$, were mapped onto certain spin models and symmetry
relations identified which allowed derivation of several
exact and asymptotic results on the
correlations and other properties
[43-49]. We note that deposition and detachment combine to
drive the dynamics into a steady state rather than jammed
state as in ordinary RSA. Since these studies are quite
recent and have been limited thus far to few $1D$ models,
we will not review them further. Multilayer deposition will
be discussed at the end of this section.

We now turn to particle motion on the surface, in a
monolayer deposit, which was experimentally observed in
deposition of proteins [16] and also in deposition on DNA
molecules [17,42]. Theoretical modeling of effects of
particle rearrangement on RSA is quite recent, and all
theoretical models reported have assumed diffusional
relaxation (random hopping in the lattice case). Consider
the dimer deposition in $1D$; see Figure~1. Hopping of
particle $b$ to the left, as indicated by an arrow,
would open up a larger gap to allow deposition
schematically marked by $c$. The configuration in Figure~1
(with particle $a$ actually deposited) is jammed in the
interval shown. Thus, diffusional relaxation allows the
deposition process to reach denser, in fact, ordered
configurations. For short times, when the empty area is
plentiful, the effect of the in-surface particle motion
will be small. However, for large times, the density will
exceed that of the RSA process, as illustrated by the
broken line in Figure~2.

Further investigation of this effect is actually simpler in
$1D$ than in $2D$. Let
us therefore consider the $1D$ case first, postponing the
discussion of $2D$ models to the next section.
Specifically, consider deposition of $k$-mers of fixed
length $V$. In order to allow limit $k \to \infty$ which
corresponds to continuum deposition, we take the
underlying lattice spacing $ \ell = V/k$. Since the
deposition attempt rate $R$ was defined per unit area (unit
length here) it has no significant $k$-dependence. However,
the added diffusional hopping of $k$-mers on the $1D$
lattice, with attempt rate $H$ and hard-core or similar
particle interaction, must be $k$-dependent. Indeed, we
consider each deposited $k$-mer particle as randomly and
independently attempting to move one lattice spacing to the
left or to the right with rate $H/2$ per unit time. Of
course, particles cannot run over each other so some sort
of hard-core interaction must be assumed, i.e., in a
dense state most hopping attempts will fail. However, if
left alone, each particle would move diffusively on large
time scales. In order to have the resulting diffusion
constant $\cal D$ finite in the continuum limit $k \to
\infty$, we put

$$ H \propto {\cal D} / \ell^2 = {\cal D}k^2 /
V^2\;\; . \eqno(3.1) $$

\noindent{}Note that (3.1) is only valid in $1D$.

Each successful hopping of a particle results in motion of
one empty lattice site (see particle $b$ in Figure~1). In
is useful to reconsider the dynamics of particle hopping in
terms of the dynamics of this rearrangement of empty area
fragments [50-52]. Indeed, if several such empty sites are
combined to form large enough voids, deposition attempts
can succeed in regions of particle density which would be
``frozen'' in ordinary RSA. In terms of these new
``particles'' which are empty lattice sites of the
deposition problem, the process is in fact that of
reaction-diffusion. Indeed, $k$ reactants (empty sites)
must be brought together (by diffusional hopping) in order
to have finite probability of their ``annihilation,'' i.e.,
disappearance of a group of consecutive nearest-neighbor
empty sites due to successful deposition. Of course, the
$k$-group can also be broken apart due to diffusion.
Therefore, the $k$-reactant annihilation is not
instantaneous in the reaction nomenclature. Such
$k$-particle reactions are of interest on their own and
they were studied by several authors [53-58].

The simplest mean-field rate equation for annihilation of
$k$ reactants describes the time dependence of the
coverage, $\rho (t)$, in terms of the reactant density
$1-\rho$,

$$ {d \rho \over dt} = \Gamma (1-\rho)^k \;\; , \eqno(3.2)
$$

\noindent{}where $\Gamma$ is the effective rate constant.
There are two problems with this approximation. Firstly,
it turns out that for $k=2$ the mean-field approach breaks
down. Diffusive-fluctuation arguments for non-mean-field
behavior have been advanced for reactions [53,55,59-60].
Actually, in $1D$, several exact calculations support this
conclusion [61-67]. Here we only note that the asymptotic
large-time behavior turns out to be

$$ 1-\rho \sim 1/\sqrt{t} \;\;\;\;\;\;\;\; (k=2,D=1) \;\; ,
\eqno(3.3) $$

\noindent{}rather than the mean-field prediction $\sim
1/t$. The coefficient in (3.3) is expected to be universal
(when expressed in an appropriate dimensionless form by
introducing single-reactant diffusion constant).
The power law (3.3) was confirmed by extensive
numerical simulations of dimer deposition [68] and by exact
solution for one particular value of $H$ [69]. The latter work
also yielded some exact results for correlations.
Specifically, while the connected
particle-particle correlations spread diffusively in space,
their decay it time is nondiffusive; see [69] for details.

The case $k=3$ is marginal with the mean-field power law
modified by logarithmic terms. The latter were not observed
in Monte Carlo studies of deposition [51]. However,
extensive results are available directly for three-body
reactions [55-58], including verification of the
logarithmic corrections to the mean-field behavior [56-58].

The second problem with the mean-field rate equation was
identified in the continuum limit of off-lattice
deposition, i.e., for $k \to \infty$. Indeed, the
mean-field approach is essentially the fast diffusion
approximation assuming that diffusional relaxation is
efficient enough to equilibrate nonuniform fluctuations on
the time scales fast as compared to the time scales of the
deposition events. Thus, the mean-field results are
formulated in terms of the uniform properties, such as
density. It turns out, however, that the simplest, $k^{\rm
th}$-power of the reactant density form (3.2) is only
appropriate for times $t >> e^{k-1}/(RV)$.

This conclusion was reached [50] by assuming the
fast-diffusion, randomized (equilibrium) hard-core reactant
system form of the inter-reactant distribution function in
$1D$ (essentially, an assumption on the form of certain
correlations). This approach, not detailed here, allows
Ginzburg-criterion-like estimation of the limits of
validity of the mean-field results and it correctly
suggests mean-field validity for $k=4,5,\ldots$, with
logarithmic violation for $k=3$ and complete breakdown of
the mean-field assumptions for $k=2$. However, this
detailed analysis yields the modified mean-field relation

$$ {d \rho \over dt } = {\gamma RV (1-\rho)^k \over \left(
1-\rho + k^{-1}\rho \right) } \;\;\;\;\;\;\;\; (D=1)
\;\; , \eqno(3.4) $$

\noindent{}where $\gamma$ is some effective dimensionless
rate constant. This new expression applies
uniformly as $k \to \infty$. Thus, the continuum deposition
is also asymptotically mean-field, with the
essentially-singular ``rate equation''

$$ {d \rho \over dt} = \gamma (1-\rho) \exp [-\rho / (1-\rho)]
\;\;\;\;\;\;\;\; (k=\infty,D=1) \;\; . \eqno(3.5) $$

\noindent{}The approach to the full, saturation coverage
for large times is extremely slow,

$$ 1 - \rho (t) \approx {1 \over \ln \left( t \ln t \right) }
\;\;\;\;\;\;\;\; (k=\infty,D=1) \;\; . \eqno(3.6) $$

\noindent{}Similar predictions for $k$-particle reactions
can be found in [55].

When particles are allowed to attach 
also on top of each other, with possibly some
rearrangement processes allowed as well, multilayer
deposits will be formed. It is important to note that the
large-layer structure of the deposit and fluctuation
properties of the growing surface will be determined by the
transport mechanism of particles to the surface and
by the allowed relaxations (rearrangements) [70-71]. Indeed, these
two characteristics determine the screening properties of
the multilayer formation process which in turn shape the
deposit morphology, which can range from fractal to dense,
and the roughening of the growing deposit surface. There is a
large body of research studying such growth, with recent
emphasis on the growing surface fluctuation properties.
However, the feature characteristic of the RSA
process, i.e., the exclusion due to particle size, plays no
role in determining the universal, large-scale properties
of ``thick'' deposits and their surfaces. Indeed, the
RSA-like jamming will be only important for detailed
morphology of the first few layers in a multilayer
deposit.

In view of the above remarks, models for which jamming has
been of interest in multilayer deposition were relatively
less studied. They can be divided into two groups. Firstly,
structure of the deposit in the first few layers is of
interest [72-74] since they retain ``memory'' of the surface.
Variation of density and other correlation properties away
from the wall has structure on the length scales of
particle size. However, these typically oscillatory
features decay away with the distance from the wall.
Secondly, few-layer deposition processes have been of
interest in some experimental systems. Mean-field
theories of multilayer deposition with particle size and
interactions accounted for were formulated [75] and used to fit
such data [11,13-15].

\

\noindent{\bf 4. Two-Dimensional Deposition with
Diffusional Relaxation}

\

We now turn to the $2D$ case of deposition of extended
objects on planar substrates, accompanied by diffusional
relaxation (assuming monolayer deposits). We note that the
available theoretical results are limited to few studies
[37,76-77]. They indicate a rich pattern of new
effects as compared to
$1D$. In fact, there exists extensive literature [78-79] on
deposition with diffusional relaxation in other models, in
particular those where the jamming effect is not present or
plays no significant role. These include, e.g.,  deposition of
``monomer'' particles which align with the underlying
lattice without jamming, as well as models where many
layers are formed (discussed in the preceding section).

As already mentioned earlier, $2D$ deposition with relaxation
of extended objects is of interest in certain experimental
systems where the depositing objects are proteins [16].
Here we focus on the combined effect of jamming and
diffusion, and we emphasize dynamics at large times [76-77].
For early stages of the deposition process, low-density
approximation schemes can be used. One such application
was reported in [37] for continuum deposition of circles
on a plane.

In order to identify features new to $2D$, let us consider
deposition of $2 \times 2$ squares on the square lattice.
The particles are exactly aligned with the $2 \times 2$
lattice sites as shown in Figure~3. Furthermore, we assume
that the diffusional hopping is along the lattice
directions $\pm x$ and $\pm y$, one lattice spacing at a
time. In this model dense configurations involve domains
of four phases as shown in Figure~3. As a result,
``immobile'' fragments of empty area can exist. Each such
single-site vacancy (Figure~3) serves as a meeting point of
four domain walls (schematically marked by the four
arrows). By ``immobile'' we mean that the vacancy cannot
move due to local motion of the surrounding particles. For
it to move, a larger empty-area fragment must first arrive,
along one of the domain walls. One such larger empty void
is shown in Figure~3. Note that it serves as a kink in the
domain wall.

Existence of immobile vacancies suggests possible
``frozen,'' glassy behavior with extremely slow relaxation,
at least locally. In fact, the full characterization of the
dynamics of this model requires further study. The first
numerical results [76] do provide many answers which,
however, will be reviewed later on. We first consider a
simpler model depicted in Figure~4. In the latter model
[77] the extended particles are smaller squares of size
$\sqrt{2} \times \sqrt{2}$. They are rotated 45$^\circ$
with respect to the underlying square lattice. Their
diffusion, however, is along the lattice axes, one lattice
spacing at a time. The equilibrium variant of this model
(without deposition, with fixed particle density) is the
well-studied hard-square model [80] which, at large
densities, phase separates into two distinct phases. These
two phases also play role in the late stages of RSA with
diffusion. Indeed, at large densities the ``immobile'' part
of the empty area is always in domain walls separating
ordered regions. One such domain wall is shown in
Figure~4. Snapshots of actual Monte Carlo simulation
results can be found in [77].

Figure~4 illustrates the process of ordering which
essentially amounts to shortening of domain walls. In
Figure~4, the domain wall gets shorter after the shaded
particles diffusively rearrange to open up a deposition
slot which can be covered by an arriving particle.
Numerical simulations [77] find behavior reminiscent of the
low-temperature equilibrium ordering processes [81-83] driven
by diffusive evolution of the domain-wall structure. For
instance, the remaining uncovered area vanishes according
to

$$ 1 - \rho (t) \sim { 1 \over \sqrt{t} } \;\; . \eqno(4.1) $$

\noindent{}This quantity, however, also measures the length
of domain walls in the system (at large times). Thus,
disregarding finite-size effects and assuming that the
domain walls are not too convoluted (as confirmed by
numerical simulations), we conclude that the power law
(4.1) corresponds to typical domain sizes growing as $\sim
\sqrt{t}$, reminiscent of the equilibrium ordering
processes of systems with nonconserved order parameter
[81-83].

We now turn to the $2 \times 2$ model of Figure~3. The
equilibrium variant of this model corresponds to
hard-squares with both nearest and next-nearest neighbor
exclusion [80,84-85]. It has been studied in lesser detail than
the simple two-phase hard-square model described in the
preceding paragraphs. In fact, the equilibrium phase
transition has not been fully classified (while it was
Ising for the simpler model). The ordering at low
temperatures and high densities was studied in [84].
However, many features noted, for instance large entropy of
the ordered arrangements, require further study. The
dynamical variant (RSA with diffusion) of this model was
studied numerically in [76]. The structure of the
single-site frozen vacancies and associated network of
domain walls turns out to be boundary-condition sensitive.
For periodic boundary conditions the density ``freezes'' at
values $1-\rho \sim L^{-1}$, where $L$ is the linear system
size.

Preliminary indications were found [76] that the domain
size and shape distributions in such a frozen state are
nontrivial. Extrapolation $L \to \infty$ indicates that
the power law behavior similar to (4.1) is nondiffusive:
the exponent $1/2$ is replaced by $\sim 0.57$. However,
the density of the smallest mobile vacancies, i.e., dimer kinks
in domain walls, one of which is illustrated in
Figure~3, does decrease diffusively. Further studies are
needed to fully clarify the ordering process associated
with approach to the full coverage as $t \to \infty$ and $L
\to \infty$ in this model.

In summary, we reviewed the deposition processes involving
extended objects, with jamming and its interplay with
diffusional relaxation yielding interesting new dynamics of
approach to the large-time state. While significant progress
has been achieved in $1D$, the $2D$ systems require further
investigations. Mean-field and low-density approximations
can be used in many instances for large enough dimensions,
for short times, and for particle sizes larger than few
lattice units. Added diffusion allows formation of denser
deposits and leads to power-law large-time tails which,
in $1D$, were related to diffusion-limited reactions, while
in $2D$, associated with evolution of domain-wall
network and defects, reminiscent of equilibrium ordering
processes. New results are likely to come from
extensive numerical simulations of both lattice and
continuum models.

\vfil\eject

\centerline{\bf REFERENCES}

\ 

{\frenchspacing

\item{[1]} Review: V. Privman, {\sl Dynamics of Nonequilibrium
Processes: Surface Adsorption, Reaction-Diffusion Kinetics, Ordering
and Phase Separation}, in {\sl Trends in Statistical
Physics}, in print (Council for Scientific Information,
Trivandrum, India).

\item{[2]} E.R. Cohen and H. Reiss,
J. Chem. Phys. {\bf 38}, 680 (1963).

\item{[3]} J. Feder and I. Giaever,
J. Colloid Interface Sci. {\bf 78}, 144 (1980).

\item{[4]} A. Schmitt, R. Varoqui, S. Uniyal, J.L. Brash and C. Pusiner,
J. Colloid Interface Sci. {\bf 92}, 25 (1983).

\item{[5]} G.Y. Onoda and E.G. Liniger,
Phys. Rev. A{\bf 33}, 715 (1986).

\item{[6]} N. Kallay, B. Bi\v skup, M. Tomi\'c and E. Matijevi\'c,
J. Colloid Interface Sci. {\bf 114}, 357 (1986).

\item{[7]} N. Kallay, M. Tomi\'c, B. Bi\v skup,
I. Kunja\v si\'c and E. Matijevi\'c,
Colloids Surfaces {\bf 28}, 185 (1987).

\item{[8]} J.D. Aptel, J.C. Voegel and A. Schmitt,
Colloids Surfaces {\bf 29}, 359 (1988).

\item{[9]} Z. Adamczyk,
Colloids and Surfaces {\bf 35}, 283 (1989).

\item{[10]} Z. Adamczyk,
Colloids and Surfaces {\bf 39}, 1 (1989).

\item{[11]} C.R. O'Melia, in {\sl Aquatic Chemical Kinetics},
p. 447, W. Stumm, ed. (Wiley, New York, 1990).

\item{[12]} Z. Adamczyk, M. Zembala, B. Siwek and P. Warszy{\' n}ski,
J. Colloid Interface Sci. {\bf 140}, 123 (1990).

\item{[13]} N. Ryde, N. Kallay and E. Matijevi\'c,
J. Chem. Soc. Farad. Tran. {\bf 87}, 1377 (1991).

\item{[14]} N. Ryde, H. Kihira and E. Matijevi\'c,
J. Colloid Interface Sci. {\bf 151}, 421 (1992).

\item{[15]} L. Song and M. Elimelech,
Colloids and Surfaces A{\bf 73}, 49 (1993).

\item{[16]} Review of deposition of proteins: J.J. Ramsden,
J. Statist. Phys. {\bf 73}, 853 (1993).

\item{[17]} Review of deposition on DNA: C.J. Murphy, M.R. Arkin,
Y. Jenkins, N.D. Ghatlia, S.H. Bossmann, N.J. Turro and J.K. Barton,
Science {\bf 262}, 1025 (1993).

\item{[18]} V.M. Glazov, S.N. Chizhevskaya and N.N. Glagoleva,
{\sl Liquid Semiconductors\/} (Plenum, New York, 1969).

\item{[19]} P. Schaaf, A. Johner and J. Talbot,
Phys. Rev. Lett. {\bf 66}, 1603 (1991).

\item{[20]} Review: M.C. Bartelt and V. Privman,
Internat. J. Mod. Phys. B{\bf 5}, 2883 (1991).

\item{[21]} Review: J.W. Evans,
Rev. Mod. Phys. {\bf 65}, 1281 (1993).

\item{[22]} B. Widom, J. Chem. Phys. {\bf 44}, 3888 (1966).

\item{[23]} B. Widom, J. Chem. Phys. {\bf 58}, 4043 (1973).

\item{[24]} P. Schaaf and J. Talbot,
Phys. Rev. Lett. {\bf 62}, 175 (1989).

\item{[25]} R. Dickman, J.-S. Wang and I. Jensen, 
J. Chem. Phys. {\bf 94}, 8252 (1991).

\item{[26]} J.J. Gonzalez, P.C. Hemmer and J.S. H{\o}ye,
Chem. Phys. {\bf 3}, 228 (1974).

\item{[27]} J.W. Evans, J. Phys. A{\bf 23}, 2227 (1990).

\item{[28]} J. Feder, J. Theor. Biology {\bf 87}, 237 (1980).

\item{[29]} E.M. Tory, W.S. Jodrey and D.K. Pickard,
J. Theor. Biology {\bf 102}, 439 (1983).

\item{[30]} E.L. Hinrichsen, J. Feder and T. J\o ssang, 
J. Statist. Phys. {\bf 44}, 793 (1986). 

\item{[31]} E. Burgos and H. Bonadeo, J. Phys. A{\bf 20}, 1193 (1987).

\item{[32]} G.C. Barker and M.J. Grimson,
J. Phys. A{\bf 20}, 2225 (1987).

\item{[33]} R.D. Vigil and R.M. Ziff,
J. Chem. Phys. {\bf 91}, 2599 (1989).

\item{[34]} J. Talbot, G. Tarjus and P. Schaaf,
Phys. Rev. A{\bf 40}, 4808 (1989).

\item{[35]} R.D. Vigil and R.M. Ziff,
J. Chem. Phys. {\bf 93}, 8270 (1990).

\item{[36]} J.D. Sherwood, J. Phys. A{\bf 23}, 2827 (1990).

\item{[37]} G. Tarjus, P. Schaaf and J. Talbot,
J. Chem. Phys. {\bf 93}, 8352 (1990).

\item{[38]} B.J. Brosilow, R.M. Ziff and R.D. Vigil,
Phys. Rev. A{\bf 43}, 631 (1991).

\item{[39]} V. Privman, J.-S. Wang and P. Nielaba,
Phys. Rev. B{\bf 43}, 3366 (1991).

\item{[40]} Y. Pomeau,
J. Phys. A{\bf 13}, L193 (1980).

\item{[41]} R.H. Swendsen,
Phys. Rev. A{\bf 24}, 504 (1981).

\item{[42]} L.S. Schulman, S.H. Bossmann and N.J. Turro,
J. Phys. Chem., in print.

\item{[43]} M. Barma, M.D. Grynberg and R.B. Stinchcombe,
Phys. Rev. Lett. {\bf 70}, 1033 (1993).

\item{[44]} R.B. Stinchcombe, M.D. Grynberg and M. Barma,
Phys. Rev. E{\bf 47}, 4018 (1993).

\item{[45]} M.D. Grynberg, T.J. Newman and R.B. Stinchcombe,
Phys. Rev. E{\bf 50}, 957 (1994).

\item{[46]} M.D. Grynberg and R.B. Stinchcombe,
Phys. Rev. E{\bf 49}, R23 (1994).

\item{[47]} G.M. Sch\"utz, J. Statist. Phys., in print.

\item{[48]} P.L. Krapivsky and E. Ben-Naim,
J. Chem. Phys. {\bf 100}, 6778 (1994). 

\item{[49]} M. Barma and D. Dhar,
Phys. Rev. Lett. {\bf 73}, 2135 (1994).

\item{[50]} V. Privman and M. Barma,
J. Chem. Phys. {\bf 97}, 6714 (1992).

\item{[51]} P. Nielaba and V. Privman,
Mod. Phys. Lett. B {\bf 6}, 533 (1992).

\item{[52]} B. Bonnier and J. McCabe, preprint {\sl Comment on
Diffusional Relaxation in $1d$ Dimer Deposition}.

\item{[53]} K. Kang, P. Meakin, J.H. Oh and S. Redner,
J. Phys. A {\bf 17}, L665 (1984).

\item{[54]} S. Cornell, M. Droz and B. Chopard,
Phys. Rev. A {\bf 44}, 4826 (1991).

\item{[55]} V. Privman and M.D. Grynberg,
J. Phys. A {\bf 25}, 6575 (1992).

\item{[56]} D. ben-Avraham,
Phys. Rev. Lett. {\bf 71}, 3733 (1993).

\item{[57]} P.L. Krapivsky,
Phys. Rev. E {\bf 49}, 3223 (1994).

\item{[58]} B.P. Lee,
J. Phys. A {\bf 27}, 2533 (1994).

\item{[59]} K. Kang and S. Redner,
Phys. Rev. Lett. {\bf 52}, 955 (1984).

\item{[60]} K. Kang and S. Redner,
Phys. Rev. A{\bf 32}, 435 (1985).

\item{[61]} Z. Racz,
Phys. Rev. Lett. {\bf 55}, 1707 (1985).

\item{[62]} M. Bramson and J.L. Lebowitz,
Phys. Rev. Lett. {\bf 61}, 2397 (1988).

\item{[63]} D.J. Balding and N.J.B. Green,
Phys. Rev. A {\bf 40}, 4585 (1989).

\item{[64]} J.G. Amar and F. Family,
Phys. Rev. A {\bf 41}, 3258 (1990).

\item{[65]} D. ben-Avraham, M.A. Burschka and C.R. Doering,
J. Statist. Phys. {\bf 60}, 695 (1990).

\item{[66]} M. Bramson and J.L. Lebowitz,
J. Statist. Phys. {\bf 62}, 297 (1991).

\item{[67]} V. Privman,
J. Statist. Phys. {\bf 69}, 629 (1992).

\item{[68]} V. Privman and P. Nielaba,
Europhys. Lett. {\bf 18}, 673 (1992).

\item{[69]} M.D. Grynberg and R.B. Stinchcombe, 
Phys. Rev. Lett. {\bf 74}, 1242 (1995).

\item{[70]} {\sl Solids Far from Equilibrium: Growth, Morphology,
Defects}, C. Godr\`eche, ed. (Cambridge University
Press, Cambridge, 1991).

\item{[71]} {\sl Dynamics of Fractal Surfaces},
F. Family and T. Vicsek, eds. (World Scientific,
Singapore, 1991).

\item{[72]} R.-F. Xiao, J.I.D. Alexander and F. Rosenberger,
Phys. Rev. A{\bf 45}, R571 (1992), and references therein.

\item{[73]} B.D. Lubachevsky, V. Privman and S.C. Roy,
Phys. Rev. E{\bf 47}, 48 (1993).

\item{[74]} Numerical Monte Carlo simulation aspects
of continuum multilayer
deposition (ballistic deposition of $3D$ balls)
were reviewed in detail
by B.D. Lubachevsky, V. Privman and S.C. Roy,
preprint {\sl Casting Pearls Ballistically: Efficient Massively
Parallel Simulation of Particle Deposition\/}.

\item{[75]} V. Privman, H.L. Frisch, N. Ryde and E. Matijevi\'c,
J. Chem. Soc. Farad. Tran. {\bf 87}, 1371 (1991).

\item{[76]} J.-S. Wang, P. Nielaba and V. Privman,
Physica A{\bf 199}, 527 (1993).

\item{[77]} J.-S. Wang, P. Nielaba and V. Privman,
Mod. Phys. Lett. B{\bf 7}, 189 (1993).

\item{[78]} Review: J.A. Vernables, G.D.T. Spiller and M. Hanb\"ucken,
Rept. Prog. Phys. {\bf 47}, 399 (1984).

\item{[79]} For more recent work, see, e.g.,
M.C. Bartelt, M.C. Tringides and J.W. Evans,
Phys. Rev. B{\bf 47}, 13891 (1993), and references therein.

\item{[80]} L.K. Runnels, in {\sl Phase Transitions and Critical
Phenomena}, Vol. 2, p. 305, C. Domb and M.S. Green, eds.
(Academic, London, 1972).

\item{[81]} J.D. Gunton, M. San Miguel, P.S. Sahni,
{\sl Phase Transitions and Critical
Phenomena}, Vol. 8, p. 267, C. Domb and J.L. Lebowitz, eds.
(Academic, London, 1983).

\item{[82]} O.G. Mouritsen, in {\sl Kinetics and
Ordering and Growth at Surfaces}, p. 1, M.G. Lagally, ed. (Plenum,
NY, 1990).

\item{[83]} A. Sadiq and K. Binder, J. Statist. Phys. {\bf 35}, 517
(1984), and references therein.

\item{[84]} K. Binder and D.P. Landau,
Phys. Rev. B{\bf 21}, 1941 (1980).

\item{[85]} W. Kinzel and M. Schick,
Phys. Rev. B{\bf 24}, 324 (1981).

}

\vfil\eject

\noindent{\bf Figure Captions}

\

\noindent\hang{}Figure 1: Deposition of dimers of the $1D$
lattice. Once the arriving dimer $a$ attaches to the
``surface,'' the configuration shown will be fully jammed
in the interval displayed. Further deposition can only
proceed if particle diffusion is allowed, as illustrated by
$b$ (and the arriving particle $c$), or by detachment of
whole units. Detachment of ``recombined'' particles to open
up landing sites, such as $d$, was also considered (see
text).

\

\noindent\hang{}Figure 2: Schematic variation of the
coverage fraction $\rho (t)$ with time for
lattice deposition without
(solid line) and with (dashed line) diffusional relaxation.
Note that the short-time behavior deviates from linear at
times of order $1/(RV)$. (Quantities $R, V, \ell$ are
defined in the text.)

\

\noindent\hang{}Figure 3: Fragment of a large-density
deposit configuration in the deposition of $2\times 2$
squares. Illustrated are one single-site ``frozen'' vacancy
at which four defect lines converge (indicated by arrows),
as well as one dimer
vacancy which causes kink in one of the domain walls
(schematically indicated by the ``kinked'' arrow).

\

\noindent\hang{}Figure 4: Illustration of deposition of
$\sqrt{2} \times \sqrt{2}$ particles on the square lattice,
fragment of which is shown separately. Diffusional motion
during time interval from $t_1$ to $t_2$ can rearrange the
empty area ``stored'' in the domain wall to open up a new
landing site for deposition. This is illustrated by the
shaded particles.

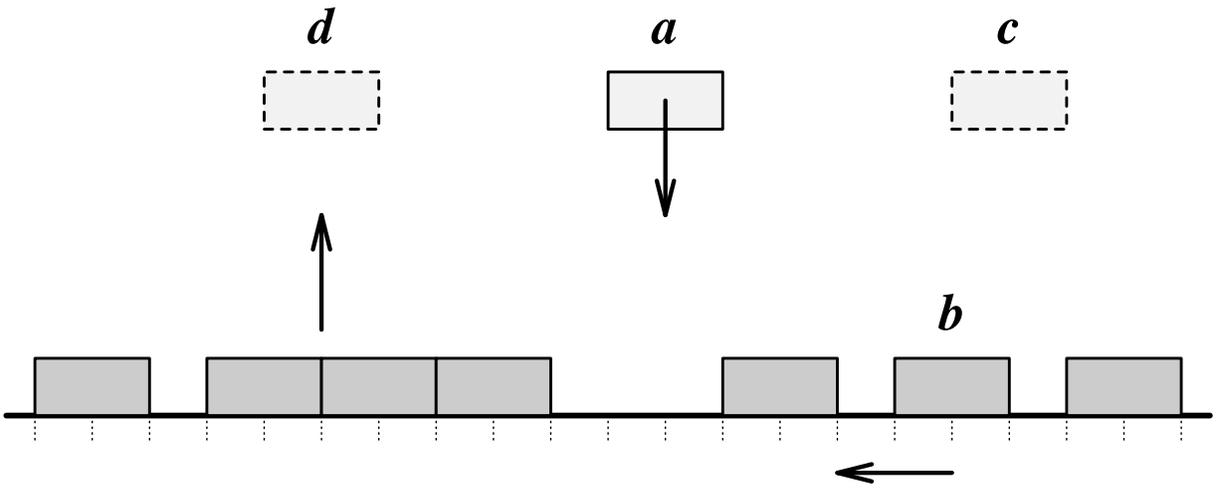
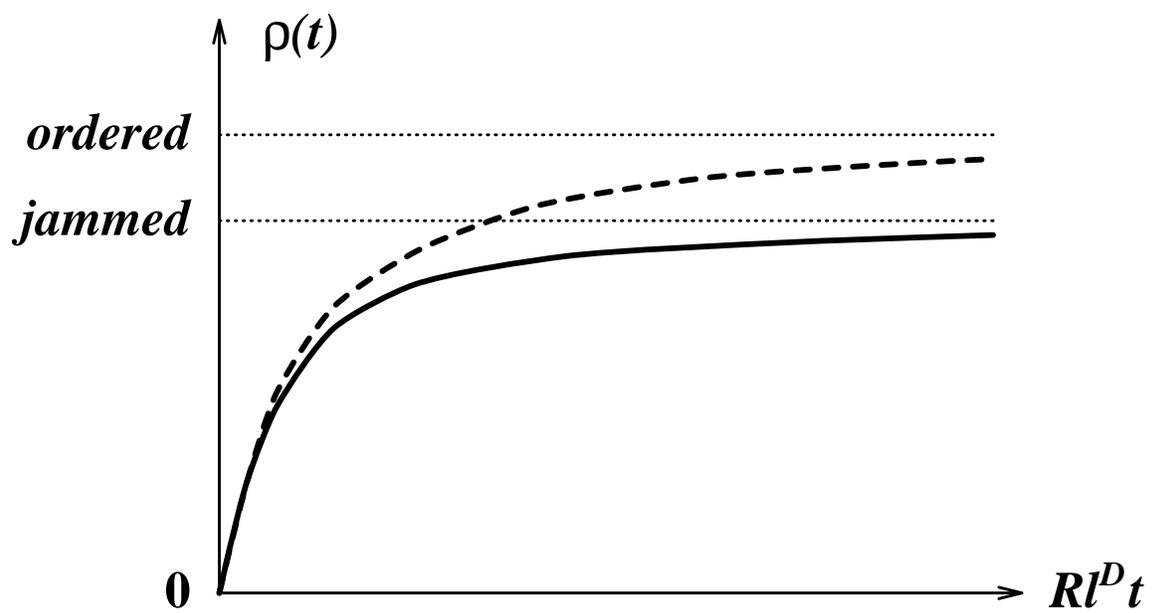
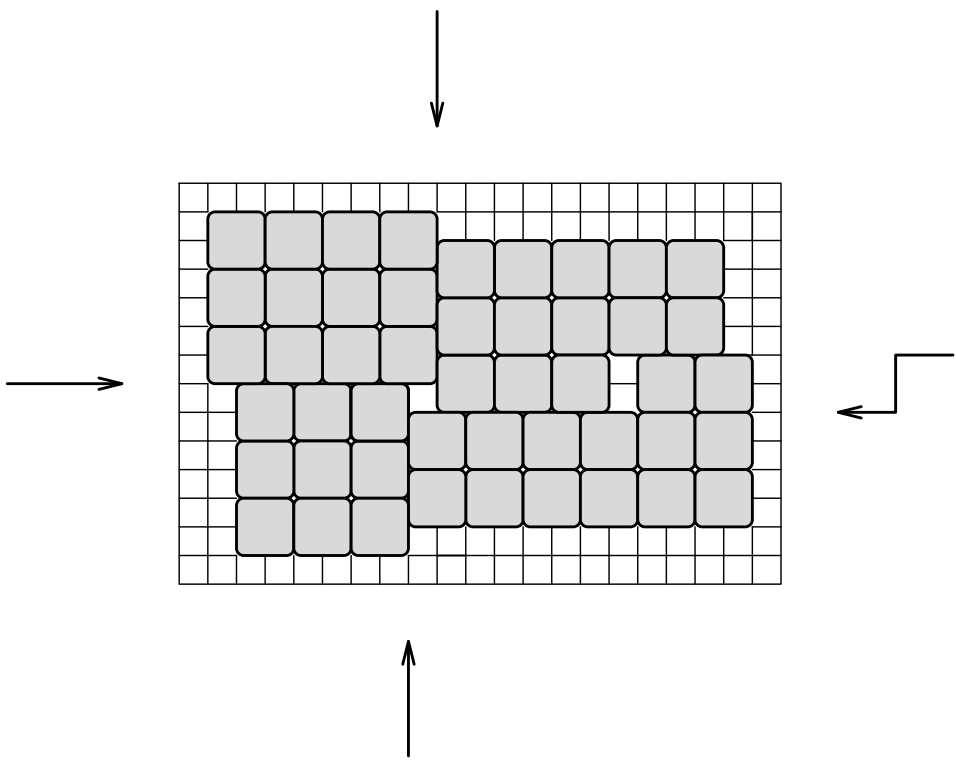
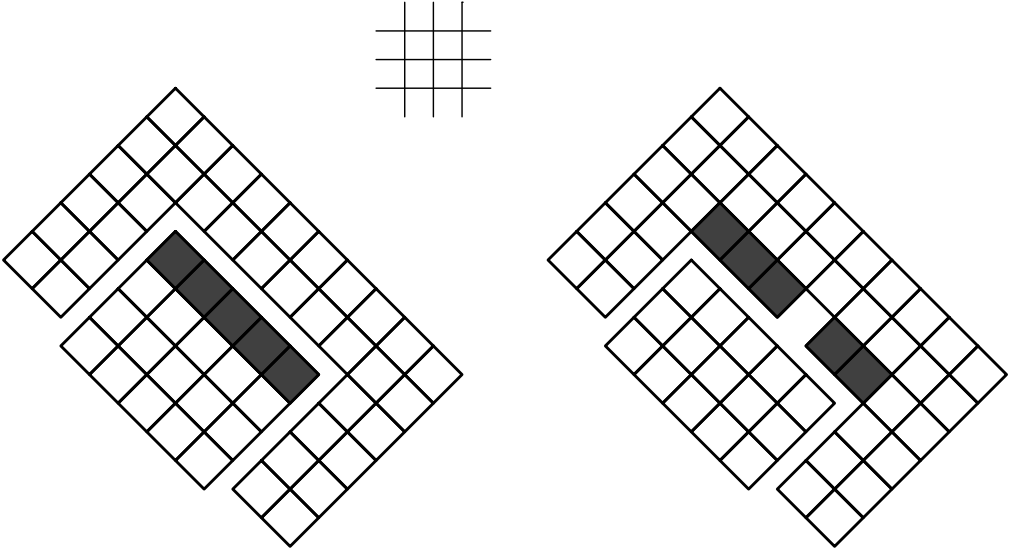
\bye